\documentclass{cernyrep} 
\usepackage{amsmath,amsfonts}
\usepackage{multirow}
\usepackage{pstricks}
\usepackage{epsfig}
\usepackage{amssymb}

\def\tv#1{\vrule height #1pt depth 5pt width 0pt}

\def\BR{{\rm{BR}}}

\begin{document}

\title{\large \bf Higgs Physics}

\author{C.~Grojean}
\institute{DESY, 22607 Hamburg, Germany\\
Institut f{\"u}r Physik, Humboldt-Universit{\"a}t zu Berlin, 12489 Berlin, Germany\\
ICREA, 08010 Barcelona, Spain (on leave)\\
IFAE, BIST, 08193 Bellaterra, Barcelona, Spain (on leave)}
\maketitle
\begin{abstract}
The cause of the screening of the weak interactions at long distances puzzled the high-energy community for more nearly half a century. With the discovery of the Higgs boson a new era started with direct experimental information on the physics behind the breaking of the electroweak symmetry. This breaking plays a fundamental role in our understanding of particle physics and sits at the high-energy frontier beyond which we expect new physics that supersedes the Standard Model. The Higgs boson (inclusive and differential) production and decay rates offer a new way to probe this frontier. 
\end{abstract}

Exhaustive reviews on electroweak symmetry breaking (EWSB) and Higgs physics are easily accessible online~\cite{Djouadi:2005gi, Reina:2012fs, Dittmaier:2012nh, Pomarol:2014dya, Csaki:2015hcd, PDG-Higgs:2016, deFlorian:2016spz} and the purpose of these notes is not to duplicate them but rather to serve as a concise introduction to the topic and to present (a personalized hence biased selection of) recent developments in the field.

\section{Introduction}
\label{sec:intro}

The Standard Model  (SM) is a triumph of the combination of the two pillars of twentieth-century physics, namely quantum mechanics and special relativity.
Particles are defined as representations of the Poincar\'e group.
Mathematically, these representations are labelled by two quantities: the spin that is quantized 
and takes only discrete values, and the mass, which a priori is a continuous parameter. 
However, the transformation laws for the various elementary particles under the 
gauge symmetries associated to the fundamental interactions force the masses 
of these particles to vanish. This would be in flagrant contradiction with the experimental measurements. 

The Brout--Englert--Higgs mechanism (BEH)~\cite{Englert:1964et,Higgs:1964ia,Higgs:1964pj,Guralnik:1964eu}
provides the solution to this mass conundrum. 
The discovery of a Higgs boson in July 2012 and the experimental confirmation of the 
BEH mechanism
by the ATLAS
and CMS
collaborations~\cite{Aad:2012tfa, Chatrchyan:2012ufa} has been a historical step 
in our understanding of nature: the masses of the elementary particles are 
not fundamental parameters defined at very high energy but rather emergent 
quantities appearing at low energy as a result of the particular structure of the vacuum.

\section{The HEP landscape after the Higgs discovery}
\label{sec:HEPlandscape}

During its first run, the LHC certainly fulfilled its commitments: 
The machine and its detectors were mostly designed to find the Higgs boson 
and ``[they] got it!" according to the words of R.~Heuer, director general 
of CERN, on 4 July 2012. It was an important step in the understanding of the mechanism of electroweak symmetry breaking. But the journey is not over.

One can ask how the Higgs discovery reshaped the High Energy Physics (HEP) landscape.
The days of theoretically guaranteed discoveries imposed on us by some no-lose theorems are over: indeed, with the addition of a light Higgs boson with a mass around 125\,GeV, the Standard Model is theoretically consistent and can be extrapolated up to very high energy, maybe as high as $10^{14\div16}$\,GeV or even the Planck scale. But at the same time, the big questions of our field, or the ones that we have considered so far as the big questions, remain wildly open: the hierarchy problem, the origin of flavor, the issue of the neutrino masses and mixings, the question of the identity of Dark Matter, the source of dynamical preponderance of matter over antimatter during the cosmological evolution of our Universe\ldots are left unanswered (see the BSM lectures~\cite{Rosenfeld:CLASHEP} in these proceedings). In the next decades, future progress in HEP is in the hands of experimentalists whose discoveries will  reveal the way Nature has solved these big questions, forcing the theorists to renounce/review/question deeply rooted bias/prejudice. The Higgs discovery sets a large part of the agenda for the theoretical and experimental HEP programs over the next couple of decades.

\section{Open questions about the Higgs}
\label{sec:HiggsQuestions}

The LHC accumulated striking evidence that the Higgs vacuum expectation value (vev) is the cause of the 
screening of the weak interaction at long distances and the source of the gauge boson masses.

However, this evidence only addresses the question of \emph{how} the symmetry of the weak
interaction is broken.  It does not address the question of
\emph{why} the symmetry is broken or why the Higgs field acquires an
expectation value. The situation is simply summarized in the following tautology
\vspace{2mm}\\
\centerline{$
\begin{array}{c}
\textrm{Why is electroweak symmetry broken?}\\
\textit{Because the Higgs potential is unstable at the origin.}\\
\textrm{Why is the Higgs potential unstable at the origin?}\\
\textit{Because otherwise EW symmetry would not be broken.}\\
\end{array}
$}
\vspace{2mm}

The discovery of a Higgs boson allowed first glimpses into a new 
sector of the microscopic world.
Now comes the time of the detailed exploration of this new Higgs sector.
And some key questions about the Higgs boson emerge:
\begin{enumerate}
\item Is it the SM Higgs?
\item Is it an elementary or a composite particle?
\item Is it unique and solitary? Or are there additional states populating the Higgs sector? 
\item Is it eternal or only temporarily living in a metastable vacuum?
\item Is its mass natural following the criteria of Dirac, Wilson or 't Hoft? 
\item Is it the first superparticle ever observed?
\item Is it really responsible for the masses of all the elementary particles?
\item Is it mainly produced by top quarks or by new heavy vector-like particles?
\item Is it a portal to a hidden world forming the dark matter component of the Universe?
\item Is it at the origin of the matter-antimatter asymmetry?
\item Has it driven the primordial inflationary expansion of the Universe?
\end{enumerate}

The answers to these questions will have profound implications on our understanding of the fundamental laws of physics.
Establishing that the Higgs boson is weakly coupled, elementary and solitary, 
would surely be as shocking as unexpected, but it may well indicate the existence of a 
multiverse ruled by anthropic selection rules.
If instead deviations from the SM emerge in the dynamics of the Higgs, we will 
have to use them as a diagnostic tool of the underlying dynamics. The pattern 
of these deviations will carry indirect information about the nature of the completion of 
the SM at higher energies. In supersymmetric models, and more generally in 
models with an extended electroweak symmetry breaking sector, the largest deviations 
will be observed in the couplings to leptons and to the down-type quarks, as well as in 
the decay amplitudes to photons and gluons. 
In models of strong interactions, in which the Higgs boson is a bound state, the effects of compositeness uniformly suppress all the Higgs couplings while the self-interactions of the particles inside the Higgs sector, namely the Higgs particle and the longitudinal 
components of the W and Z bosons, will increase with the transferred energy.
Moreover, the measurements of the Higgs couplings will also reveal the 
symmetry properties of the ``Higgs boson" observed.
For instance, it can be established whether the new scalar is indeed ``a Higgs" fitting into a $\textrm{SU}(2)$ doublet together with the degrees of freedom associated with 
the longitudinal W and Z and not some exotic impostor, like for instance a pseudo-dilaton.  If the Higgs is found to have an internal structure, a detailed study of the Higgs couplings can also establish whether it is just an ordinary composite, like a $\sigma$ particle, or whether it is a pseudo-Nambu--Goldstone boson endowed with additional symmetry properties, 
like the $\pi$'s of QCD.

\section{What is the SM Higgs the name of?}
\label{sec:SMHiggs}

\subsection{The SM Higgs boson as a UV regulator}

The SM Higgs boson ensures the proper decoupling of the longitudinal polarizations of the massive EW gauge bosons at high energy.
Indeed, these
longitudinal modes of  W$^\pm$ and $Z$ can
be described by Nambu--Goldstone bosons associated
to the coset $\textrm{SU}(2)_{\rm L}\times \textrm{SU}(2)_{\rm R} / \textrm{SU}(2)_{\mbox{\scriptsize
    isospin}}$. Their kinetic term corresponds to the gauge boson mass terms,
\begin{eqnarray}
\frac{1}{2}m_Z^2 Z_\mu Z^\mu + m_W^2 W_{\mu}^+ W^{-\mu}=
\frac{v^2}{4} \mbox{Tr} ( D_\mu \Sigma^\dagger D^\mu \Sigma)
\label{eq:kin}
\end{eqnarray}
with $\Sigma = e^{i\sigma^a \pi^a/v}$, where $\sigma^a$ ($a=1,2,3$) are the usual Pauli matrices. Due to the
Goldstone boson equivalence theorem the non-trivial scattering of the 
longitudinal gauge bosons $V$ ($V=W^\pm,Z$) is controlled by the contact
interactions among four pions from the expansion of the Lagrangian
of Eq.~(\ref{eq:kin}), leading to amplitudes growing with the energy,
\begin{equation}
{\cal A} (V^a_L V^b_L \to V^c_L V^d_L) = {\cal A}(s) \delta^{ab}
\delta^{cd} + {\cal A} (t) \delta^{ac} \delta^{bd} 
 + {\cal A} (u)
\delta^{ad} \delta^{bc} \quad \mbox{with} \quad {\cal A} (s) \approx
\frac{s}{v^2} \:.
\end{equation}
Here $s,t,u$ denote the Mandelstam variables, and $v$ represents the ``Higgs vev", with $v=246$~GeV. The amplitude grows with the
center-of-mass (c.m.) energy squared $s$, and therefore perturbative unitarity
will be lost around $4 \pi v \sim 3$~TeV, unless there is a new weakly coupled
elementary degree of freedom. The simplest realization of new dynamics
restoring perturbative unitarity is given by a single scalar field
$h$, which is singlet under $\textrm{SU}(2)_{\rm L}\times \textrm{SU}(2)_{\rm R} / \textrm{SU}(2)_{\mbox{\scriptsize
    isospin}}$ and
couples to the longitudinal gauge bosons and fermions
as~\cite{Giudice:2007fh,Contino:2010mh}, 
\begin{eqnarray}
\label{eq:lewsb} 
&&{\cal L}_{\rm EWSB} = \frac{1}{2} (\partial_\mu h)^2 - V(h) +
\frac{v^2}{4} \,\mbox{Tr} (D_\mu \Sigma^\dagger D^\mu \Sigma) \left(
  1+2a \frac{h}{v} + \sum_{n\geq 2} b_n \frac{h^n}{v^n}+\ldots
  \right) \nonumber \\
&&\hspace*{1.3cm} - \frac{v}{\sqrt{2}}
(\bar{u}^i_L \bar{d}^i_L) \Sigma \left( 1 + c \frac{h}{v} + \sum_{n\geq 2}c_n
  \frac{h^n}{v^n} + ... \right) \left(\begin{array}{c} y^u_{ij} u^j_R
    \\ y^d_{ij} d^j_R \end{array}\right)    +{\rm h.c.}\ \nonumber
\end{eqnarray}
with
\begin{eqnarray}
\hspace{-.7cm}
V(h) = \frac{1}{2} m_h^2 h^2 + \frac{d_3}{6} \left(\frac{3m_h^2}{v}\right) h^3 + \frac{d_4}{24}
\left(\frac{3m_h^2}{v^2}\right) h^4 + ... 
\label{eq:vewsb}
\end{eqnarray}
For $a=1$ the scalar exchange cancels the piece growing with the
energy in the $V_L V_L$ amplitude. If in addition $b_2=a^2$ then also in
the inelastic amplitude $V_L V_L \to hh$ perturbative unitarity is maintained,
while for $ac=1$ the $V_L V_L \to f f'$ amplitude also remains
finite. The SM Higgs boson is defined by the point $a=b_2=c=1$ and
$d_3=d_4=1$, $c_{n\geq2}=b_{n\geq3}=0$. The scalar resonance and the pions then
combine to from a doublet which transforms linearly under $\textrm{SU}(2)_{\rm L}
\times \textrm{SU}(2)_{\rm R}$. 

\begin{figure}[!ht]
\centerline{\includegraphics[width=0.7\hsize]{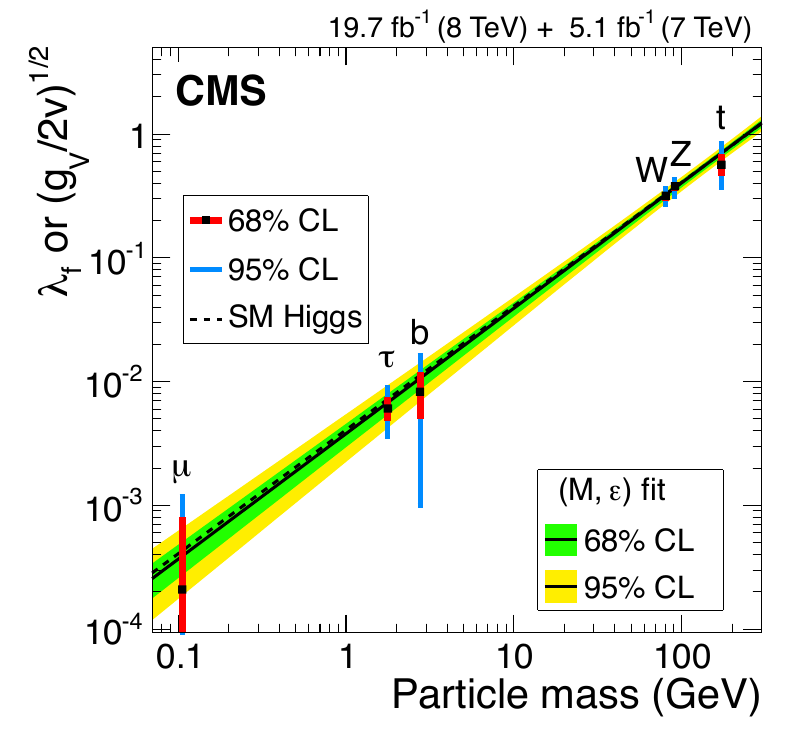}}
\caption[]{Comparison of the SM predictions (black dashed line) and the fit of the LHC measurements of the Higgs couplings to various SM particles. From Ref.~\cite{Khachatryan:2014jba}.}
\label{fig:Higgsfits}
\end{figure}

The requirement that the Higgs boson fully unitarizes the scattering amplitudes of massive particles therefore implies that the Higgs couplings to the various SM particles are directly proportional to their masses. This fundamental property is in remarkable agreement with the state-of-the-art fit of the current Higgs data collected at the LHC, see Fig.~\ref{fig:Higgsfits}.

The couplings to the heaviest particles, namely $W$ and $Z$ bosons, the top quark and the $\tau$ lepton, are already established. The measurement of the couplings to other quarks and leptons, in particular the lightest ones, will require considerably more statistics at 
the LHC. 
Nonetheless, it is already established that the Higgs boson, contrary to all the gauge 
bosons, has non-universal couplings among the particles of the three different 
generations of quarks and leptons. The Higgs particle is not a $Z'$ gauge boson! The Higgs boson mediates new fundamental forces different in nature than  the electromagnetic, weak and strong interactions. Are other forces of this type going to be discovered? Models of DM and baryogenesis make use of new forces like the ones mediated by the Higgs boson.

\subsection{The flavor preserving nature of the SM Higgs}

In the SM, the Yukawa interactions are the only source of the fermion masses and they also generate linear interaction with the physical Higgs boson
\begin{eqnarray}
Y_{ij}\, \bar{\psi}_i H \psi_j = \frac{Y_{ij} v}{\sqrt{2}} \bar{\psi}_i \psi_j + \frac{Y_{ij}}{\sqrt{2}} h\, \bar{\psi}_i \psi_j.
\end{eqnarray}
Clearly both matrices can be diagonalized simultaneously and this ensures the absence of flavor changing neutral currents induced by the Higgs boson exchange. 

This nice property is no longer true if the SM fermions mix with vector-like partners or in the presence of generic higher dimension Yukawa interactions (see for instance Ref.~\cite{Buras:2011ph} for a general phenomenological discussion):
\begin{equation}
Y_{ij} \left(1+ \frac{c_{ij}}{f^2} |H|^2 \right) \bar{\psi}_i H \psi_j = \frac{Y_{ij} v}{\sqrt{2}} \left(1+ \frac{c_{ij}v^2}{2 f^2} \right)  \bar{\psi}_i \psi_j 
+ \frac{Y_{ij}}{\sqrt{2}} \left(1+ \frac{3 c_{ij} v^2}{2f^2} \right) h\, \bar{\psi}_i \psi_j.
\end{equation}

Therefore it is particularly important to probe the flavor structure of the Higgs interactions and to look for flavor-violating decays, e.g. $h \to \mu \tau$, or production, e.g. $t \to h c$. Limits from low-energy flavor-changing interactions indirectly constrain these processes especially in the quark sector but leave the possibility of sizeable effects in the lepton sector  (see for instance Ref.~\cite{Harnik:2012pb} for an extensive discussion). The slight 2.5$\sigma$ excess in the $h\to \mu \tau$ decay initially reported by CMS with the full run~1 dataset~\cite{Khachatryan:2015kon} is  confirmed neither by the CMS run~2 data~\cite{CMS:2016qvi}, nor by the ATLAS run~1 analysis~\cite{Aad:2016blu}. Nonetheless, these analyses start probing the interesting region of parameter space where the off-diagonal Yukawa couplings are set by the mass of the leptons, $|Y_{\mu\tau} Y_{\tau\mu}| \sim m_\tau m_\mu/v^2$, one order of magnitude better than the indirect bounds set by $\tau \to \mu \gamma$ and $\tau \to 3\mu$.

\section{The SM Higgs boson at the LHC}
\label{sec:SM-Higgs@LHC}

\begin{figure}[!ht]
\centerline{\includegraphics[width=0.60\hsize]{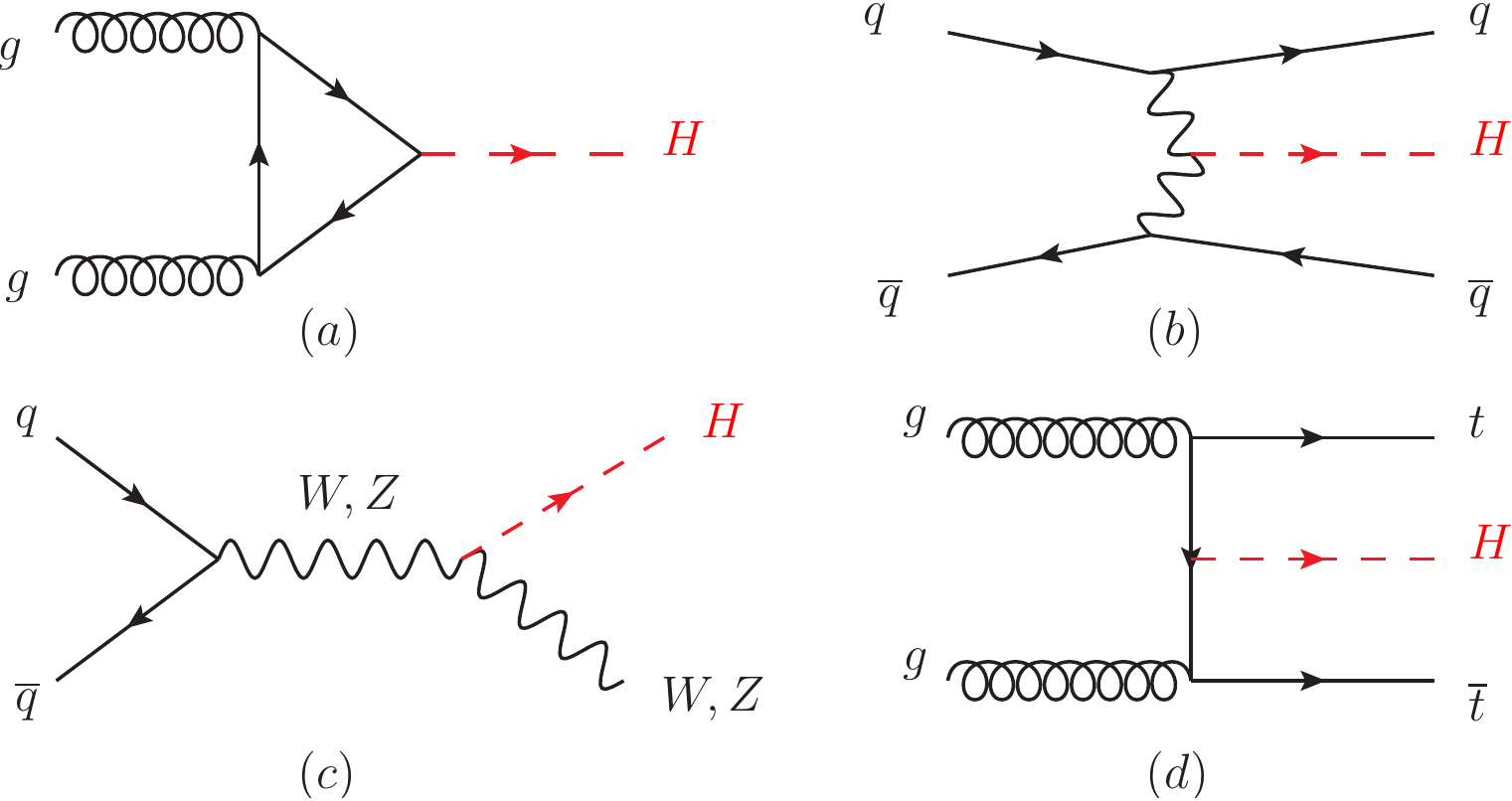}}
\caption[]{Generic Feynman diagrams contributing to the Higgs production 
in (a) gluon fusion, (b) weak-boson fusion, (c) Higgs-strahlung (or associated production with a gauge boson) and  (d) associated production
 with top quarks. From Ref.~\cite{PDG-Higgs:2016}.}
\label{fig:FeynmanHprod}
\end{figure}

The main production mechanisms at the LHC are gluon fusion, weak-boson fusion, associated production with a gauge boson and associated production
with a pair of top/antitop quarks.
Figure~\ref{fig:FeynmanHprod} depicts representative diagrams for these dominant Higgs production processes. 

\begin{figure}[!ht]
\centerline{\includegraphics[width=0.45\hsize]{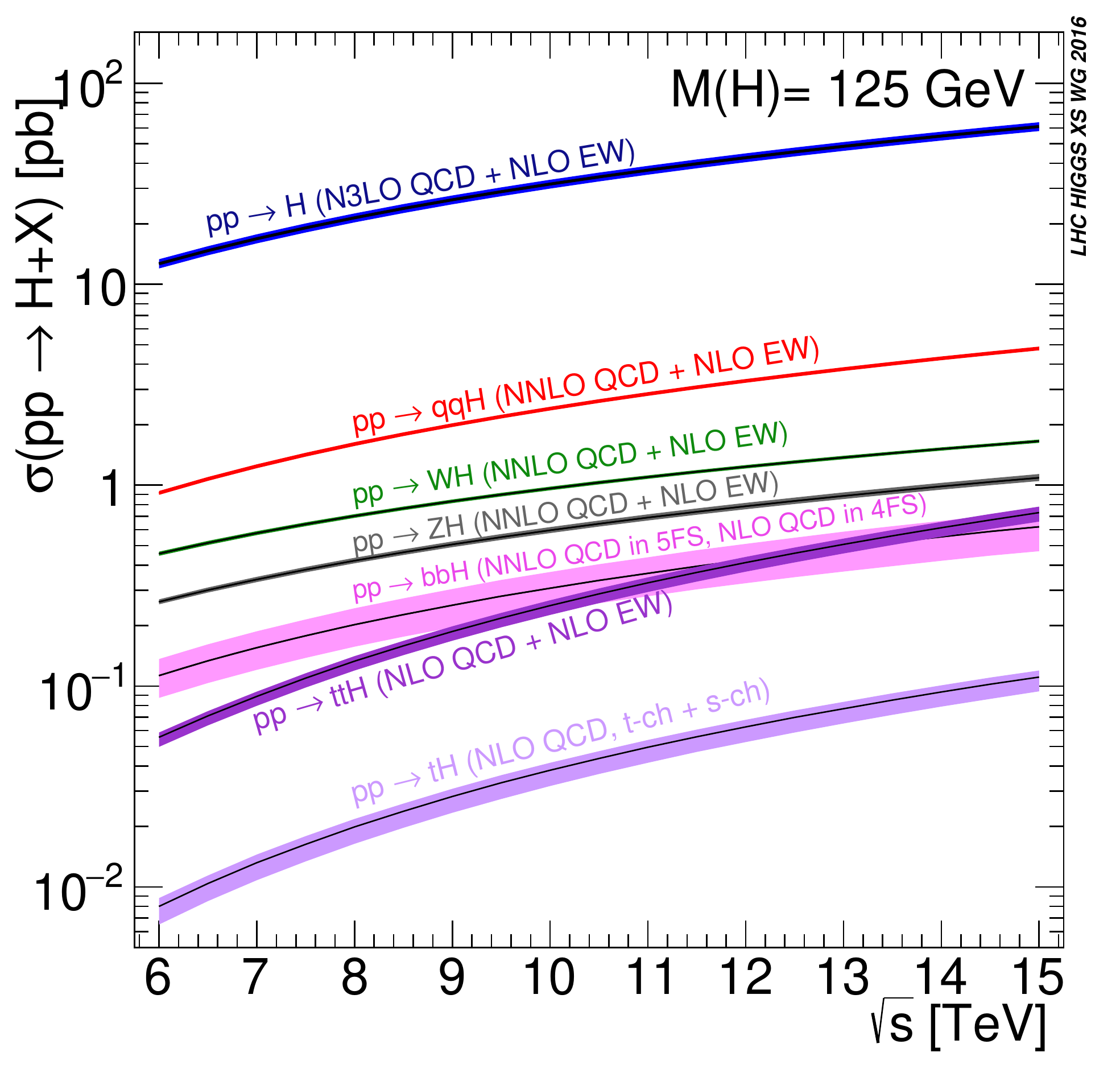}
\includegraphics[width=0.45\hsize]{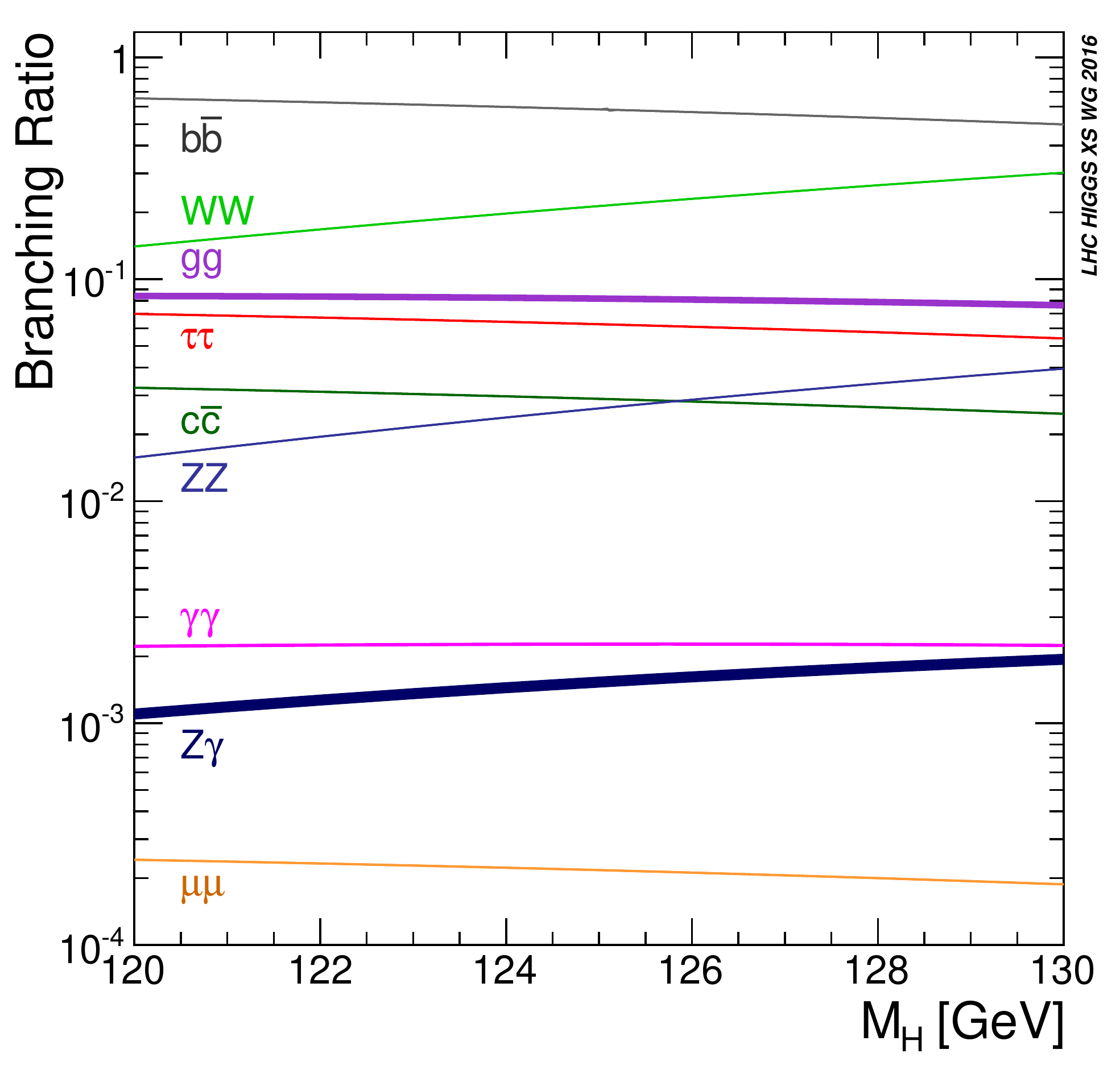}}
\caption[]{(Left) The SM Higgs boson production cross sections as a function of
the center of mass energy, $\sqrt{s}$, for $pp$ collisions.  (Right) The branching ratios for the main decays of the SM
Higgs boson near $m_H=125$\,GeV. For both plots, the theoretical
uncertainties are indicated as bands. From Ref.~\cite{PDG-Higgs:2016}.}
\label{fig:lhcxs}
\end{figure}

The cross sections for the production of a SM Higgs boson  as a function of $\sqrt{s}$, the center of mass energy, for $pp$ collisions, are
summarized 
in Figure~\ref{fig:lhcxs}(left).
A  detailed discussion, including  uncertainties in the 
theoretical calculations due to missing higher-order 
effects and experimental uncertainties on the determination 
of SM parameters involved in the calculations can be found 
in Refs.~\cite{Dittmaier:2011ti, Dittmaier:2012vm, Heinemeyer:2013tqa, deFlorian:2016spz}. 
These references also discuss the impact of PDF's uncertainties, QCD scale uncertainties and uncertainties due to different matching procedures  when including higher-order corrections matched to parton shower simulations as well as uncertainties due to hadronization and parton-shower events. 
Table~\ref{table:StateArt} summarizes state-of-the-art of the theoretical calculations in the main different production channels.

\begin{table*}[!ht]
\begin{center}
\begin{tabular}{cccc}
\hline
ggF &  VBF & VH &  $t{\bar t}H$ \\
\hline
Fixed order:  & Fixed order: & Fixed order: & Fixed order: \\
NNLO QCD + NLO EW & NNLO QCD & NLO QCD+EW& NLO QCD\\
{\tt \small (HIGLU, iHixs, FeHiPro, HNNLO)} & {\tt (VBF@NNLO)} & {\tt (V2HV and HAWK)}& {\tt (Powheg)} \\
Resummed: & Fixed order: & Fixed order: & {\tt (MG5\_aMC@NLO)}\\
{\small NNLO + NNLL QCD} & {\small NLO QCD + NLO EW} & {\small NNLO QCD} &\\
{\tt (HRes)} & {\tt (HAWK)} & {\tt (VH@NNLO)} & \\
Higgs $p_T$: & & & \\
NNLO+NNLL & & & \\
{\tt (HqT, HRes)} & & & \\
Jet Veto: & & & \\
N3LO+NNLL  & & &
\\\hline
 \end{tabular}
\vspace{.4cm}
\caption[]{\label{table:StateArt}
State-of-the-art of the theoretical calculations in the main different Higgs production channels in the SM, and main MC tools used in the simulations. From Ref.~\cite{PDG-Higgs:2016}.}
\end{center}
\end{table*}

Among other subdominant production channels of the Higgs boson at the LHC, the production in association with a single top quark, the production in association with a pair of bottom quarks and the production in association with a pair of charm quarks are particularly interesting and may become visible with the high statistics of the HL-LHC run.

Figure~\ref{fig:lhcxs} (right) reports the SM predictions for the decay fractions of the Higgs boson. A Higgs mass of about 125\,GeV provides  an excellent opportunity to explore the Higgs couplings to many SM particles. In particular  the dominant decay modes  are $H \rightarrow b\bar b$ and  $H \rightarrow WW^*$,  
followed by $H \rightarrow gg $, $H \rightarrow \tau^+\tau^-$,  $H \rightarrow c\bar c$ and $H \rightarrow Z Z^*$. 
With much smaller rates follow the Higgs decays into $H \rightarrow \gamma \gamma$, $H \rightarrow \gamma Z$ and $H \rightarrow \mu^+\mu^-$. 
Since the decays into gluons, diphotons and $Z \gamma$ are loop induced, they provide indirect information on the Higgs couplings to $WW$, $ZZ$ and $t{\bar t}$ in different combinations.
The uncertainties in the branching ratios include the missing higher-order corrections in the theoretical calculations as well as the errors in the SM input parameters, in particular fermion masses and the QCD gauge coupling, involved in the decay. The state-of-the-art  calculations of the theoretical uncertainties  is discussed in Ref.~\cite{deFlorian:2016spz}.

\section{The Higgs mass as a model-discriminator}
\label{sec:mass}

\begin{figure}[!ht]
\centerline{\includegraphics[width=0.85\hsize]{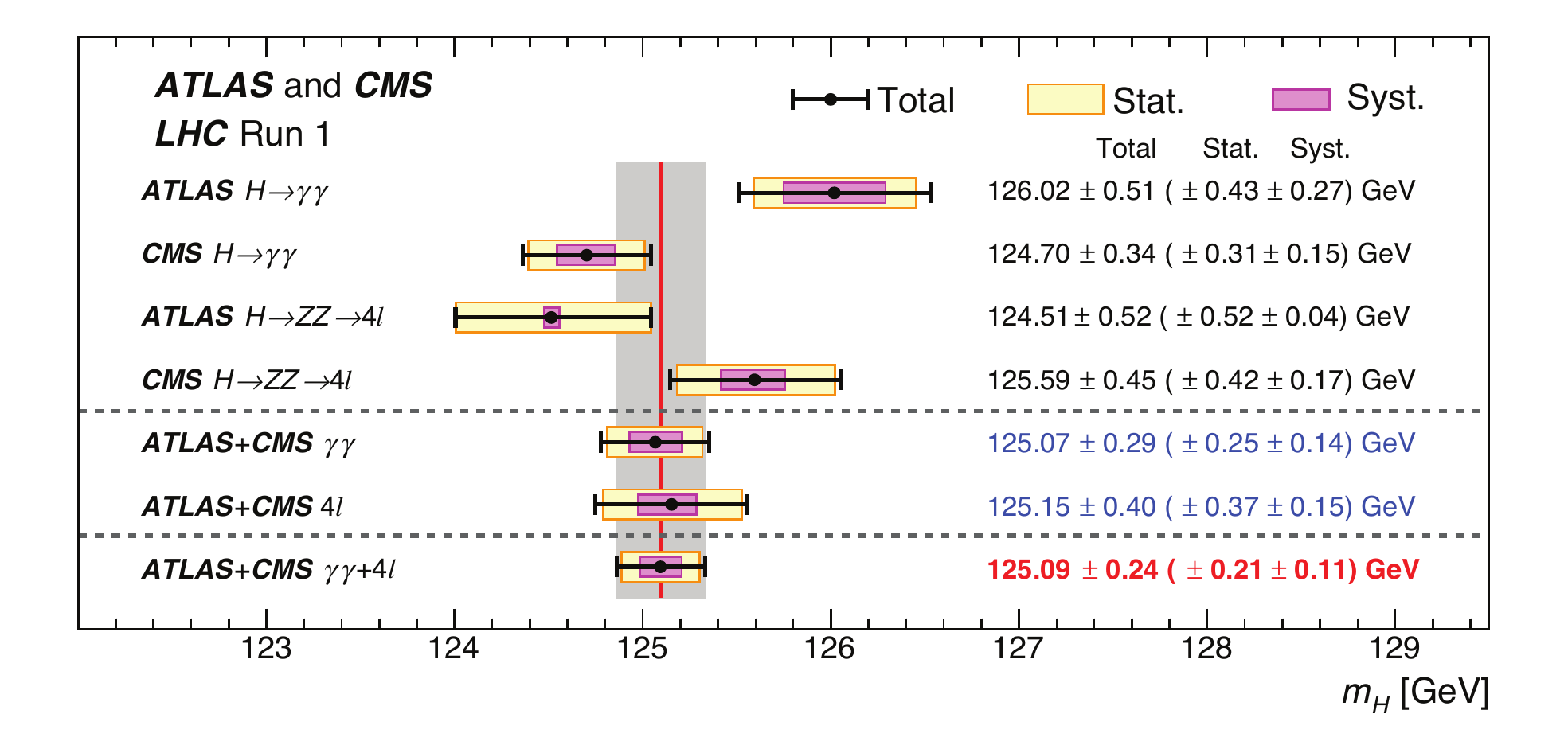}}
\caption[]{Compilation of the CMS and ATLAS mass measurements in the  $\gamma\gamma$ and $ZZ$ channels, the combined result from each experiment and  their combination.  From Ref.~\cite{Aad:2015zhl}.}
\label{fig:MH}
\end{figure}

As indicated in the previous section, the value of the Higgs boson mass opens many decay modes at a rate accessible experimentally. 
Two channels are particularly accurate in accessing the Higgs mass: $H\to \gamma\gamma$ and $H \to ZZ^* \to 2\ell^+2\ell^-$. 
Figure~\ref{fig:MH} summarizes the mass measurements in these two channels and their
combination~\cite{Aad:2015zhl}. The ATLAS and CMS combined mass
measurement:
$$ m_H = 125.09 \pm 0.21 ({\rm stat.})  \pm 0.11 ({\rm syst.})\,{\rm GeV} $$ 
reaches a precision of 0.2\% and is dominated by statistical uncertainties.

Under the assumption that the SM laws govern Nature up to very high energy,  the precise value of the Higgs mass has thrilling implications on the stability of the EW vacuum and hence the fate of our Universe (see for instance Ref.~\cite{Buttazzo:2013uya} for an extensive list of references).

The value of the Higgs mass also gives clues about the details of possible Ultra-Violet  (UV) completions of the SM itself.
This can be exemplified in the leading scenarios, namely the Minimal Supersymmetric Model (MSSM) and the Minimal Composite Higgs model (MCHM). 
In short, the Higgs mass is larger than what is typically expected in the MSSM and smaller than what is expected in the MCHM. At the classical/Born level, the mass of the lightest MSSM (SM-like) Higgs boson is bounded to be lower than the Z-boson mass since supersymmetry dictates the Higgs quartic to be fixed in terms of the gauge couplings. Some significant amount of radiative corrections, mostly from the top and stop sectors, are therefore called to raise the value of the Higgs mass. At one-loop, the Higgs mass can be approximated by
\begin{equation}
M_h^2  \simeq M_Z^2 \cos^2 2 \beta  +\frac{3 \sqrt{2} G_F M_t^4}{16 \pi^2} 
\left[\log \frac{M_{\tilde t}^2}{M_t^2} + \frac{X_t^2}{M_{\tilde t}^2} \left(1-\frac{X_t^2}{12 M_{\tilde t}^2} \right) \right]\ ,
\end{equation}
where $M_{\tilde t}^2 = M_{Q_3} M_{U_3}$ is the geometric mean of the stop masses and $X_t$ is the mixing between the two stops. Clearly, a Higgs boson as heavy as 125\,GeV requires either heavy stops ($M_{\tilde t}>800$\,GeV) and/or maximally mixed stops ($X_t \simeq \sqrt{6} M_{\tilde t}$), which brings back some amount of irreducible fine-tuning or call for non-trivial boundary conditions like non-universal gaugino masses at high-energy. Going beyond the minimal model, for instance by adding an extra gauge singlet, can easily help increasing the Higgs mass with significantly less amount of tuning, see for instance Ref.~\cite{Farina:2013fsa} for a discussion.
 
In the Minimal Composite Higgs models, the Higgs boson emerges from the strong sector as a pseudo-Nambu--Goldstone boson.  Therefore, the strong interactions themselves are not responsible for generating a potential for the Higgs field, that is generated only at the one-loop level from the interactions between the strong sector and the SM.
Computing the details of the potential from first principles remains out of reach but it is possible~\cite{Panico:2015jxa} to estimate the Higgs mass using general properties about the asymptotic behavior of correlators, i.e. imposing the saturation of the Weinberg sum rules with the first few light resonances, to obtain
\begin{equation}
M_h^2 \simeq \frac{3 M_t^2 M_Q^2}{\pi^2 f^2}\, ,
\end{equation}
where $f$ is the scale of the strong interactions (the decay constant of the Higgs boson, the equivalent of $f_\pi$ for the QCD pions) and $M_Q$ is the typical mass scale of the fermion resonances (aka the top partners). This estimate can read as
\begin{equation}
\hspace{-.6cm}
M_Q \simeq 700\,\textrm{GeV} \left( \frac{M_h}{125\,{\rm GeV}} \right) \left( \frac{160\,{\rm GeV}}{M_t} \right) \left( \frac{f}{500\,{\rm GeV}} \right).
\end{equation}
For a natural set-up ($v^2/f^2 \leq 0.2$), we therefore expect some light top partners below one TeV.
The discovery of such fermionic top-partners would be a first evidence of a strong dynamics at the origin of the breaking of the electroweak symmetry. 
\section{The Higgs profile as a probe of new physics}
\label{sec:Higgscouplings}

A dedicated study of the Higgs boson properties and couplings offers a way to infer what the structure of physics beyond the Standard Model can be.
Natural models trying to give a rationale for why/how the Higgs mass is screened from high energy corrections at the quantum level generically predict some deviations in the Higgs couplings compared to the SM predictions of the order 1\% to 100\%.
The current Higgs data accumulated at the LHC by the ATLAS and CMS collaborations already constrain the Higgs couplings to massive gauge bosons and to fermions not to deviate by more than 20--30\% from the SM predictions~\cite{Khachatryan:2016vau}.

In general, new physics can deform the SM in many ways but most of these deformations are already severely constrained by electroweak precision measurements or flavor data. Assuming flavor universality among the couplings between the Higgs boson and the SM fermions, it was shown~\cite{Gupta:2014rxa,Pomarol:2014dya} that  eight directions among the leading CP-conserving deformations of the SM  can be probed, at tree-level, only in processes with a physical Higgs boson.
These deformation induce deviations in the Higgs couplings that respect the Lorentz structure of the SM interactions, or generate simple new interactions of the Higgs boson to the $W$ and $Z$ field strengths,  
 or induce some contact interactions of the Higgs boson to photons (and to a photon and a Z boson) and gluons that take the form of the ones that are generated by integrating out the top quark. In other words, the Higgs couplings are described, in the unitary gauge, by the following effective Lagrangian~\cite{LHCHiggsCrossSectionWorkingGroup:2012nn,Heinemeyer:2013tqa}
\begin{eqnarray}
\label{L_kappa}
&& \hspace{-1.2cm}
\mathcal{L}  
= \kappa_3 \frac{m_H^2}{2 v} H^3+ 
\kappa_Z \frac{m_Z^2}{v} Z_{\mu}Z^{\mu} H 
+ \kappa_W \frac{2 m_W^2}{v} W^+_{\mu} W^{-\mu} H   +\kappa_g \frac{\alpha_s}{12 \pi v} G^a_{\mu \nu} G^{a\mu \nu} H 
\nonumber\\[.1cm]
&& \hspace{-.8cm}
+ \kappa_{\gamma} \frac{\alpha}{2\pi v} A_{\mu \nu} A^{\mu \nu} H + \kappa_{Z\gamma} \frac{\alpha}{\pi v} A_{\mu \nu} Z^{\mu \nu} H+ \kappa_{VV} \frac{\alpha}{2\pi v} \left(  \cos^2 \theta_W Z_{\mu\nu}Z^{\mu\nu}  + 2\, W_{\mu\nu}^+ W^{-\mu\nu}\right) H \cr
&& \hspace{-.8cm}
 -\left( \kappa_t \sum_{f=u,c,t} \frac{m_f}{v} \bar{f}_L  f_R
+\kappa_b \sum_{f=d,s,b} \frac{m_f}{v} \bar{f}_L  f_R +\kappa_\tau \sum_{f=e,\mu,\tau} \frac{m_f}{v} \bar{f}_L  f_R +h.c. \right) H.
\end{eqnarray}
In the SM, the Higgs boson does not couple to massless gauge bosons at tree level, hence $\kappa_{g}=\kappa_\gamma=\kappa_{Z\gamma}=0$.
Nonetheless, the contact operators are generated radiatively by loops of SM particles. 
In particular, the top quark gives a contribution to the 3 coefficients $\kappa_{g}, \kappa_\gamma, \kappa_{Z\gamma}$ that does not decouple in the infinite top mass limit. For instance, in that limit $\kappa_\gamma=\kappa_g=1$~\cite{Ellis:1975ap,Shifman:1979eb}.

The coefficient for the contact interactions of the Higgs boson to the $W$ and $Z$ field strengths is not independent but obeys the relation
\begin{equation}
(1-\cos^4 \theta_W) \kappa_{VV} = \sin 2\theta_W\, \kappa_{Z\gamma} + \sin^2 \theta_W\,  \kappa_{\gamma}.
\label{custodial kappa's}
\end{equation}
This relation is a general consequence of the so-called custodial symmetry~\cite{Contino:2013kra}. When the Higgs boson is part of an SU(2)$_{\rm L}$ doublet, the custodial symmetry could only be broken by a single operator at the level of dimension-6 operators and it is accidentally realized among the interactions with four derivatives, like the contact interactions considered. Custodial symmetry also implies 
$\kappa_Z=\kappa_W,$
leaving exactly 8 free couplings~\cite{Gupta:2014rxa,Pomarol:2014dya}. Out of these 8 coefficients, only $\kappa_V$ can be indirectly constrained by EW precision data at a level comparable from the direct constraints from LHC Higgs data~\cite{Ciuchini:2013pca}.

\begin{figure}[!ht]
\centerline{\includegraphics[width=0.8\hsize]{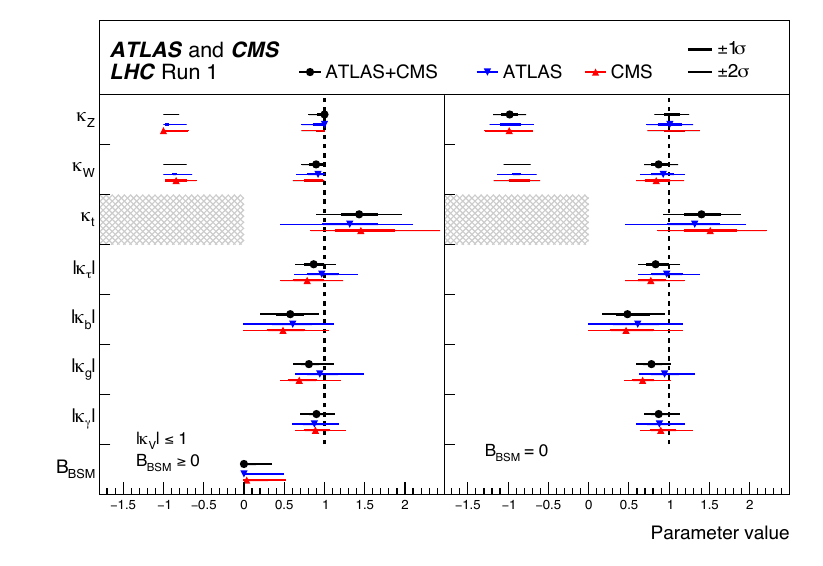}}
\caption[]{\label{fig:FullKappa}
ATLAS-CMS combined measurements of coupling modifiers. From Ref.~\cite{PDG-Higgs:2016}.}
\end{figure}

Table~\ref{table:muSummary} reports the best measurements of the production cross section times branching ratio for the main Higgs channels.
These measurements constitute a stress-test of the SM itself (any deviation from $\mu_i=1$ being an indication of new physics) but they are also used as inputs to fit the $\kappa$ coupling modifiers.
Under several assumptions, for instance on the total width of the Higgs boson, a global fit, as the one reported on Fig.~\ref{fig:FullKappa}, can be performed.

\begin{table}[!ht]
\begin{center}
\begin{tabular}{lrrrrrr}
\hline
\tv{12}               & $\gamma\gamma$            & ZZ (4$\ell$)             & WW ($\ell\nu\ell\nu$)     &  $\tau^+\tau^-$          & $b\overline{b}$   & Comb.             \\
\hline
\tv{14} ggF  &$1.10^{+0.22}_{-0.21}{}^{+0.07}_{-0.05}$&$1.13^{+0.33}_{-0.30}{}^{+0.09}_{-0.07}$&  $0.84^{+0.12}_{-0.12}{}^{+0.12}_{-0.11}$&$1.00^{+0.4}_{-0.4}{}^{+0.4}_{-0.4}$&---&$1.03^{+0.16}_{-0.14}$\\
\tv{14}VBF  &$1.3\pm 0.5{}^{+0.2}_{-0.1}$&$0.1^{+1.1}_{-0.6}{}^{+0.2}_{-0.2}$&$1.2^{+0.4}_{-0.3}{}^{+0.2}_{-0.2}$&$1.3^{+0.3}_{-0.3}{}^{+0.2}_{-0.2}$&---&$1.18^{+0.25}_{-0.23}$\\
\tv{14}WH &$0.5^{+1.3}_{-1.2}{}^{+0.2}_{-0.1}$&---&$1.6^{+1.0}_{-0.9}{}^{+0.6}_{-0.5}$&$-1.4^{+1.2}_{-1.1}{}^{+0.7}_{-0.8}$&$1.0^{+0.4}_{-0.4}{}^{+0.3}_{-0.3}$&$0.89^{+0.40}_{-0.38}$\\
\tv{14}ZH  &$0.5^{3.0}_{-2.5}{}^{+0.5}_{-0.2}$&---&$5.9^{+2.3}_{-2.1}{}^{+1.1}_{-0.8}$&$2.2^{+2.2}_{-1.7}{}^{+0.8}_{-0.6}$&$0.4^{+0.3}_{-0.3}{}^{+0.2}_{-0.2}$&$0.79^{+0.38}_{-0.36}$\\
\tv{14}ttH &$2.2^{1.6}_{-1.3}{}^{+0.2}_{-0.1}$&---&$5.0^{+1.5}_{-1.5}{}^{+1.0}_{-0.9}$&$-1.9^{+3.2}_{-2.7}{}^{+1.9}_{-1.8}$&$1.1^{+0.5}_{-0.5}{}^{+0.8}_{-0.8}$ &  $2.3^{+0.7}_{-0.6}$\\
\hline
\tv{14}Comb. &$1.14^{+0.19}_{-0.18}$&$1.29^{+0.26}_{-0.23}$&$1.09^{+0.18}_{-0.16}$&$1.11^{+0.24}_{-0.22}$&$0.70^{+0.29}_{-0.27}$&$1.09^{+0.11}_{-0.10}$ \\
\hline
\end{tabular}
\vspace{.4cm}
\caption[]{\label{table:muSummary}
Summary of the combined measurements of the $\sigma \times \BR$ for the five main
production and five main decay modes. When uncertainties are separated
into two components, the first is the statistical uncertainty and the
second is the systematic uncertainty. When only one uncertainty is
reported, it is the total uncertainty. From Ref.~\cite{PDG-Higgs:2016}.}
\end{center}
\end{table}

The effective Lagrangian of Eq.~(\ref{L_kappa}) can be amended by 6 extra Higgs couplings that break the CP symmetry
\begin{eqnarray}
\label{L_kappa_CPodd}
&&\hspace{-1.2cm}
\mathcal{L}   =
\tilde{\kappa}_g \frac{\alpha_s}{12 \pi v} G^a_{\mu \nu} \tilde{G}^{a\mu \nu} H +
\tilde{\kappa}_{\gamma} \frac{\alpha}{2\pi v} A_{\mu \nu} \tilde{A}^{\mu \nu} H + \tilde{\kappa}_{Z\gamma} \frac{\alpha}{\pi v} A_{\mu \nu} \tilde{Z}^{\mu \nu} H \nonumber\\[.1cm]
&&\hspace{-.8cm}
 - i \left( \tilde{\kappa}_t \sum_{f=u,c,t} \frac{m_f}{v} \bar{f}_L  f_R
+ \tilde{\kappa}_b \sum_{f=d,s,b} \frac{m_f}{v} \bar{f}_L  f_R + \tilde{\kappa}_\tau \sum_{f=e,\mu,\tau} \frac{m_f}{v} \bar{f}_L  f_R +h.c. \right) H,
\end{eqnarray}
where $\tilde{F}_{\mu\nu}= \epsilon_{\mu\nu\rho\sigma} F^{\rho\sigma}$ is the dual field-strength of $F_{\mu\nu}$.   It is certainly tempting to consider  new sources of CP violation in the Higgs sector, potentially bringing in one of the necessary ingredients for a successful baryogenesis scenario. It should be noted~\cite{McKeen:2012av} that these CP violating couplings would induce quark and electron electric dipole moments at  one- (for $\tilde{\kappa}_{\gamma}$ and $\tilde{\kappa}_{Z\gamma}$) or two-loops  (for $\tilde{\kappa}_{f}$).   Unless the Yukawa couplings of the Higgs to the electron and light quarks are significantly reduced compared to their SM values,  these constraints severely limit  the possibility to observe any CP violating signal in the Higgs sector at the LHC.

The coefficient $\kappa_3$ can be accessed only through double Higgs production processes, hence it will remain largely unconstrained at the LHC and a future machine like an ILC~\cite{Fujii:2015jha} or a future very high-energy  circular collider might be needed to pin down this coupling~\cite{Contino:2016spe}.
The LHC will also have a limited sensitivity on the coefficient $\kappa_\tau$ since the lepton contribution to the Higgs production cross section remains subdominant and the only way to access the Higgs coupling is via the $H \to \tau^+ \tau^-$  and possibly $H \to \mu^+ \mu^-$ channels.
Until the associated production of a Higgs with a pair of top quarks is observed, the Higgs coupling to the top quark is only probed indirectly via the one-loop gluon fusion production or the radiative decay into two photons. However, these two processes are only sensitive to the two combinations $(\kappa_t+\kappa_g)$ and $(\kappa_t+\kappa_\gamma)$ and a deviation in the Higgs coupling to the top quark can in principle always be masked by new contact interactions to photons and gluons (for a discussion, see Ref.~\cite{Grojean:2013nya}). In the next section, we shall discuss how individual information on $\kappa_{\gamma, g, t}$ can be obtained by studying either the hard recoil of the Higgs boson against an extra jet or the off-shell Higgs production in $gg\to h^* \to ZZ \to 4\ell$.

\section{Beyond inclusive single Higgs measurements}
\label{sec:off-shell+boosted}

So far the LHC has mostly produced Higgs bosons on-shell in processes with a characteristic scale around the Higgs mass.
This gives a rather good portrait of the Higgs couplings around the weak scale itself. However to fully accomplish its role as a  UV regulator of the scattering amplitudes, what matter are the couplings of the Higgs at asymptotically large energy.  To probe the Higgs couplings at large energy, one can rely on the associated production  with additional boosted particle(s) but the price to pay is to deal with significantly lower production rates.

\subsection{Boosted Higgs}
\label{sec:boosted}

The dominant production mode of the Higgs at the LHC is the gluon fusion channel.
This is a purely radiative process. The lightness of the Higgs boson plays a malicious role and makes it impossible to disentangle short- and long-distance contribution to the total rate. This limitation is embodied in the Higgs low energy theorem~\cite{Ellis:1975ap, Shifman:1979eb} that prevents one from resolving the loop contribution itself (the NLO gluon fusion inclusive cross section for a finite and infinite  top mass differ only by 1\%, see Ref.~\cite{Grazzini:2013mca}). 
New Physics could modify the top Yukawa and also generate a contact interaction to the gluons without leaving any impact on the total rate, provided that $ \kappa_t+\kappa_g=1$. Concrete examples are top partners in composite Higgs models or mixed stops in the MSSM. Still, extra radiation in the $gg\to h$ process  will allow one to explore the structure of the top loop. When the extra radiation carries away a large amount of energy and boosts the Higgs boson, the process effectively probes the ultraviolet structure of the top loop. Notice that the extra radiation cannot be in the form of a photon, as the amplitude for $gg\to h + \gamma$ vanishes due to Furry's theorem. One is therefore led to consider the production of $h$ in association with a jet.

\begin{figure}[!ht]
\centerline{\includegraphics[width=0.80\hsize]{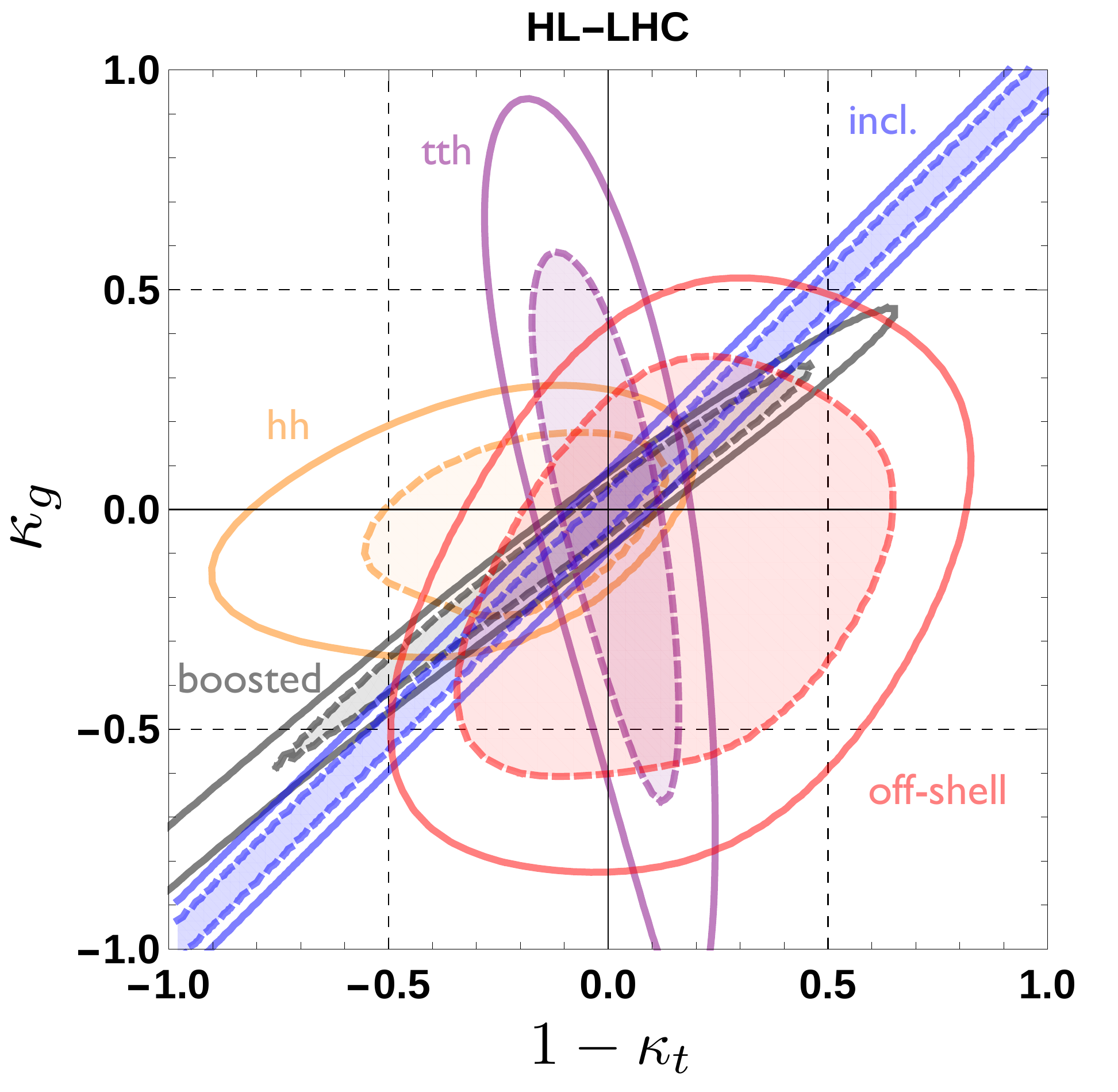}}
\caption[]{$95\%$ (solid) and $68\%$ (dashed) exclusion contours in the $(\kappa_t,\,\kappa_g)$ plane obtained from HL-LHC projections (assuming a $14$\,TeV $pp$ run with $3$\,ab$^{-1}$ of integrated luminosity): inclusive Higgs measurements (blue), $t\bar{t}h$ (purple), off-shell (red), boosted (gray), and double Higgs production (orange). From Ref.~\cite{Azatov:2016xik}.}
\label{fig:beyond-inclusive}
\end{figure}

Figure~\ref{fig:beyond-inclusive} gives the sensitivity on the boosted analysis to resolve the $\kappa_t$--$\kappa_g$ degeneracy plaguing the inclusive rate measurement~\cite{Grojean:2013nya}. Similar results have been obtained in Refs.~\cite{Banfi:2013yoa,Azatov:2013xha} and a more realistic analysis of $h\to 2\ell + \textnormal{jet}$ via $h\to \tau\tau$ and $h\to WW^*$ has been performed in Ref.~\cite{Schlaffer:2014osa}.

It should be noted that the $gg\to h+\textrm{jet}$ process has been computed only at leading order with the full mass dependence on the fermion running in the loops.  The theoretical uncertainty can be estimated by relying on the NNLO $K$-factors computed in the $m_t \to \infty$ limit. It is however clear that an exact NLO computation of the SM Higgs $p_{T}$ spectrum would be very welcome.

\subsection{Off-shell Higgs}
\label{sec:offshell}

As for any other quantum particle, the influence of the Higgs boson  is not limited to its mass shell. In 2014, the CMS and ATLAS collaborations reported the differential cross-section measurement of $pp \to Z^{(*)}Z^{(*)} \to 4\ell,2\ell2\nu$ ($\ell = e, \mu$) at high invariant-mass of the $ZZ$ system~\cite{Khachatryan:2014iha}. This process receives a sizable contribution from a Higgs produced off-shell by gluon fusion~\cite{Glover:1988rg}. 
As such, this process potentially carries information relevant for probing the EFT at large momenta and could thus reveal the energy-dependence of the Higgs couplings controlled by higher-dimensional operators with extra derivatives. It has been proposed~\cite{Caola:2013yja} to use the off-shell Higgs data to bound, in a model-independent way,  the Higgs width. However this bound actually holds under the assumption that the Higgs couplings remain unaltered over a large range of energy scales and thus applies only to very specific models. Instead, the off-shell measurement offers a rather unique access to the structure of the Higgs couplings at high energy. Again this channel reveals itself to be particularly efficient to disentangle the long and short distance contribution to the Higgs production by gluon fusion.

Figure~\ref{fig:beyond-inclusive} also shows the sensitivity on the off-shell analysis to resolve the $\kappa_t$--$\kappa_g$ degeneracy plaguing the inclusive rate measurement~\cite{Azatov:2014jga}. For a recent discussion of off-shell Higgs production within the SM and beyond and an extensive list of references, see Ref.~\cite{Kauer:2015pma}.

\section{Conclusions}
\label{sec:conclusions}

The first run of the LHC operations crowned the Standard Model as the successful description of the fundamental constituents of matter and their interactions to the tiniest details, from the QCD jet production over many orders of magnitude, to the multiple productions of electroweak gauge bosons as well as the production of top quarks. Undeniably, the Higgs boson discovery will  remain the acme of the LHC run~1 and it has profound theoretical and phenomenological implications. 
The LHC run~2 at $\sqrt{s}=13$\,TeV has already beautifully confirmed the pivotal role of the Higgs boson in the Standard Model and it is expected that on its way towards its full high-luminosity run,  the LHC will provide invaluable and crucial experimental information on the physics behind the breaking of the electroweak symmetry and it carries the hopes to finally reveal the first cracks in the SM grounds. If naturalness turned out to be a good guide, the LHC should soon find new states and revolutionize the field. If we are not so lucky and such new states are too heavy for the LHC reach, we might still detect indirectly their presence through the deviations they can induce on the Higgs properties. Precise measurements of such properties are therefore crucial and could be extremely useful to guide future direct searches at higher energies, either at the LHC itself or at other future facilities. 

The Higgs boson might also be a portal to a hidden sector whose existence is anticipated to account for the total  matter and energy budget of the Universe.  The Higgs boson could also be one key agent in driving the early exponentially growing phase of our Universe and thus allowing large scale structures to emerge from original quantum fluctuating seeds.

The search for the Higgs boson has occupied the particle physics community
for the last 50 years. With the Higgs discovery in 2012, High Energy Physics experiences a profound change in paradigm: What used to be the missing particle in the SM now quickly turns into a tool both to explore the
manifestations of the SM and to possible venture into the physics landscape beyond.
Whatever the LHC will reveal next, the exploring of new territory is on-going and for sure we, as high-energy practitioners, are living in exciting times!

\section*{Acknowledgements}
I would like to thank N.~Ellis, M.~Mulders, E.~Carrera for the organization of CLASHEP and the invitation to report on the latest developments on Higgs physics.
I also  thank all the students, and the other lecturers, for the nice atmosphere during the school.


\providecommand{\href}[2]{#2}\begingroup\raggedright

\end{document}